\documentclass[fleqn,10pt]{wlscirep}
\title{A biofilm and organomineralisation model for the growth and limiting size of ooids}

\author[1,2]{Murray T. Batchelor}
\author[3,4,*]{Robert V. Burne}
\author[5]{Bruce I. Henry}
\author[6,7]{Fei Li}
\author[8]{Josef Paul}
\affil[1]{Centre for Modern Physics, Chongqing University, Chongqing 400044, China}
\affil[2]{Mathematical Sciences Institute and Department of Theoretical Physics, Research School of Physical Sciences and Engineering, 
Australian National University, Canberra, ACT 2601, Australia}
\affil[3]{Research School of Earth Sciences, Australian National University, Canberra, ACT 2601, Australia}
\affil[4]{School of Earth and Environmental Sciences, The University of Queensland, Brisbane, QLD 4072, Australia}
\affil[5]{School of Mathematics and Statistics, UNSW Sydney, Sydney 2052, Australia}
\affil[6]{State Key Laboratory of Oil and Gas Reservoir Geology and Exploitation, Southwest Petroleum University, Chengdu 610500, China}
\affil[7]{Key Laboratory of Carbonate Reservoir of CNPC, Department of Sedimentology and Hydrocarbon Accumulation, Southwest Petroleum University, Chengdu 610500, China}
\affil[8]{Abt. Sedimentologie, Umweltgeologie, Geowissenschaftliches Zentrum der Universit\"at, Goldschmidt-Str. 3, 37077 G\"ottingen, Germany}

\affil[*]{Correspondence should be addressed to robert.burne@anu.edu.au}

\begin{abstract}
Ooids are typically spherical sediment grains characterised by concentric layers encapsulating a core. 
There is no universally accepted explanation for ooid genesis, though factors such as agitation, 
abiotic and/or microbial mineralisation and size limitation have been variously invoked. 
We develop a mathematical model for ooid growth, inspired by work on avascular brain tumours, that assumes mineralisation in a biofilm to form a central core 
and concentric growth of laminations. The model predicts a limiting size with the sequential width variation of growth rings comparing favourably with those observed 
in experimentally grown ooids generated from biomicrospheres. In reality, this model pattern may be complicated during growth by syngenetic aggrading neomorphism 
of the unstable mineral phase, followed by diagenetic recrystallisation that further complicates the structure. 
Our model provides a potential key to understanding the genetic archive preserved in the internal 
structures of naturally occurring ooids.
\end{abstract}
\begin{document}

\flushbottom
\maketitle

\thispagestyle{empty}

\section*{Introduction}

Ooids are, typically, spherical sediment grains characterised by a core encapsulated by a cortex of concentric layers (see Fig.~\ref{figvarious}). 
In some cases they are nucleated on a detrital grain or a shell fragment, but any unevenness inherited from an irregularly shaped nucleus 
is smoothed out by successive cortical layers until a spherical form is attained\cite{Flugel1978,Simone1980,Richter1983}. 
Although they have been known since ancient times\cite{B2012} there is, as yet,  no universally accepted explanation for their origin. 
Conflicting interpretations of ooid genesis include the aggregation of fine grained particles around a nucleus while rolling on a soft substrate\cite{S1879}, 
a chemical origin by precipitation from a supersaturated solution around a nucleus\cite{B1971}, 
a biochemical origin in which mineral precipitation is catalysed by dissolved organic matter\cite{SF1972}, 
and a biological origin through the organomineralisation of a surface biofilm\cite{S1970}. 
Attempts to synthesise ooids in the laboratory have led to inconclusive results\cite{SF1972}. 
Currently favoured models of ooid formation mainly emphasise the effects of rolling or agitation\cite{SG1993,Trower2017}. 
Here we examine the alternative possibility that ooid genesis is initiated by the colonisation of a spherical surface by a biofilm\cite{S2003,L2017} 
that mineralises in a similar manner to that suggested by the experiments and analysis of Brehm, Krumbein and Palinska\cite{BKP2006}. 
Recent investigations demonstrate that phototrophic, heterotrophic, aerobis and anaerobic microbes are all involved in the mineralisation of modern ooids\cite{Diaz2017}. 
We have developed this concept into a mathematical model for the influence of a biofilm on the growth of ooids.  
Our model is inspired by the principles underlying Greenspan's model of avascular brain tumours\cite{G1972}. 
The model assumes initial organomineralisation influenced by microbial organisms in a biofilm and captures the features of a central core, 
concentric growth of laminations and a limiting size. 
The sequential width variation of growth rings can be compared directly with those observed in experimentally grown ooids generated from biomicrospheres\cite{BKP2006}.

\section*{Results}

\subsection*{Mathematical model.} 

We assume that the biofilm contains microbial organisms that require nutrients to survive and to multiply\cite{S2003}.  
If the nutrient level falls below a critical concentration then the microbes cannot grow. 
It is assumed that the nutrients come from the surrounding media and diffuse into the biofilm, 
similar to the diffusion of nutrients in the tumour model of Greenspan\cite{G1972}. 
In Greenspan's model for avascular tumour growth, nutrients diffuse in towards a growing tumour and are taken up by the tumour at a constant rate per unit time. 
The key predictions that have been made in the mathematical analysis of this model are: (i) The tumor will develop a necrotic core. 
This occurs after the tumour reaches a critical size, when the nutrients are taken up in the outer parts of the tumour before they have time to diffuse into the centre of the tumour. 
The morphology of the tumour is then a solid necrotic core surrounded by an outer layer that is still receiving nutrients. 
(ii) The size of the necrotic core increases until, after long times, the width of the outer layer, and the overall size of the tumour reach constant limiting sizes. 
The width of the outer layer depends on the background nutrient level, the rate of take up of nutrients by the tumour, and the threshold nutrient level required by the tumour not to necrotise. 

In our model we hypothesise that the overall growth of ooids is similar to that of avascular tumours, where growth occurs in a nutrient limited environment. 
We suppose that ooids form in a biofilm whose growth is dependent on the take up of diffusing nutrients at a constant rate per unit time. 
Following the analysis of Greenspan, under this hypothesis, the biofilm will be characterised by two regions: 
An outer layer that is supplied by diffusing nutrients, and an inner zone that nutrients cannot diffuse into before they are taken up by the biofilm in the outer layer. 
Furthermore, similar to Greenspan's model for avascular tumour growth, the overall size of the biofilm, and the width of the active region that is supplied by nutrients, will approach constant limits.

For simplification we assume that the growth is radially symmetric. 
The growth with these model assumptions is characterised by a threshold size beyond which there will be two distinct regions, 
an outer region in which the biotic organisms grow and an inner region  in which they do not. 
In the inner region we will suppose that the microbes will either decay or become overwhelmed by mineralisation.
In the inner region there will therefore be an increase in volume due to mineralisation and a loss of volume due to decay. 
We assume these are two competing processes.
In the outer region there will be an increase in volume due to microbial colonisation.

Let $V_b^I(t)$ denote the volume of biotic organism in inner region I, and let $V_b^{II}(t)$ denote the volume
of biotic organism in outer region II.
We assume growth rates as follows:\\ 
(i) the rate of growth of the mineralisation volume is proportional to the volume of biotic organisms in the inner zone,
\begin{equation}
\frac{dV_m}{dt}=k_m V_b^I(t).
\label{eqn1}
\end{equation}
(ii) the rate of decay of biotic volume is proportional to the  volume of microbes in the inner zone,
\begin{equation}
\frac{dV_b^{I}}{dt}=-k_b^I V_b^I(t).
\label{eqn2}
\end{equation}
(iii) the rate of growth of biotic volume in the outer region is proportional to the volume of microbes in the outer zone,
\begin{equation}
\frac{dV^{II}_b}{dt}=k_b^{II} V_b^{II}(t).
\label{eqn3}
\end{equation}

\noindent

The assumption underlying equation (\ref{eqn1}) is that the organomineralisation occurs in the inner region and it is dependent on the microbial organisms in that region.   
The simplest dependence is that it is proportional to the amount of microbial organisms in that region. 
Note that equation (\ref{eqn2}) for the breakdown of microbial organisms also assumes a proportional dependence on the amount of microbial organisms in the region. 
This might be expected to be a reasonable assumption for microbes that no longer have access to nutrients. 
The constants of proportionality in equations (\ref{eqn1}) and (\ref{eqn2}) are taken to be different. 
The assumption underlying equation (\ref{eqn3}) is that there is a constant per capita growth rate of microbes if they have sufficient nutrients.
In addition to the above we assume that microbes from region II will transition into region I at a rate $k_b$ proportional to the volume of microbes in region II.
The model equations are then given by
\begin{eqnarray}
&&\frac{dV_m}{dt}=k_mV_b^I(t),\label{eqn1k}\\
&&\frac{dV_b^I}{dt}=-k_b^IV_b^I(t)+k_b V_b^{II}(t),\label{eqn2k}\\
&&\frac{dV^{II}}{dt}=k_b^{II}V_b^{II}(t)-k_b V_b^{II}(t)\label{eqn3k}.
\end{eqnarray}

Finally we assume that the volume of the outer region is proportional to the volume of microbes in the outer zone, $V_{II}(t)=c V_b^{II}(t)$. 
The volume of the inner region is the volume of mineralisation plus a volume that is proportional to the volume of microbes in the inner zone,
$V_{I}(t)=V_m(t)+cV_b^{I}(t)$.
The volume, $V(t)=V_I(t)+V_{II}(t)$, of the entire ooid is thus 
\begin{equation}
V(t)=V_m(t)+cV_b^{I}+cV_b^{II}.
\label{eqn4}
\end{equation}
We now consider predictions from the model equations (\ref{eqn1k})-(\ref{eqn4}), under the overriding hypothesis that the overall growth is limited, similar to the Greenspan model for avascular tumours. 
The meanings of variables and parameters in our model equations are summarised in Table~\ref{tabvars}.

The ooid will reach a limiting size if $\frac{dV}{dt}=0$. From the above we have 
\begin{equation}
\frac{dV}{dt}=-(ck_b^I-k_m)V_b^I(t)+ck_b^{II} V_b^{II}(t).
\end{equation}
For the limiting size we note, as in the avascular tumour model\cite{G1972},  that
\begin{equation}
cV_b^{II}=V^{II}\sim 4\pi R^2 w, 
\end{equation}
where $w$ is the constant width and 
\begin{equation}
cV_b^{I}+V_m=V^{I}\sim \frac{4}{3}\pi R^3.
\end{equation} 
The constant width depends on the threshold nutrient concentration required by microbes to survive, $c_I$, the background nutrient concentration
in the outer region, $c_{II}$, the diffusivity of the nutrients, $D$, and the rate of take up of nutrients by microbes $k$. 
In the case of spherical growth this is given by\cite{G1972} 
\begin{equation}
w=\sqrt{2(c_{II}-c_{I})\,D/k}. 
\label{width}
\end{equation}

The limiting size thus follows from the equation 
\begin{equation}
-(k_b^I-\frac{k_m}{c})(\frac{4}{3}\pi R^3-V_m)+k_b^{II}4\pi R^2 w =0 \,.
\end{equation}
It is constructive to write this equation as
\begin{equation}
-\frac{4}{3}\pi | k_b^I-\frac{k_m}{c}| R^3+4\pi k_b^{II}wR^2+|k_b^I-\frac{k_m}{c}|V_m=0 \,.
\end{equation}
This equation is of the form $-AR^3+BR^2+C=0$ with exactly one sign change in the coefficients, thus by Descartes rule of signs there is exactly one positive
root $R$. 
It is possible to write down an explicit algebraic solution for $R$ but we can get a better physical understanding of the solution by considering upper and lower bounds, 
and scaling with $w$. 
First we note that the limiting radius $R$ increases with increasing $V_m$. We can thus obtain a lower bound for $R$ by setting $V_m=0$, then
\begin{equation}
R>\frac{3ck_b^{II}w}{ck_b^I-k_m}.
\end{equation}
Note that $V_m<\frac{4}{3}\pi(R-w)^3$, so that an upper bound for $R$ can be found by setting $V_m=\frac{4}{3}\pi R^3-4\pi R^2w+4\pi Rw^2$, and then
\begin{equation}
R< \frac{w(ck_b-k_m)}{ck_b^I-k_m-ck_b^{II}}.
\end{equation}
Thus
\begin{equation}
\frac{3ck_b^{II}w}{ck_b^I-k_m}<R_{max}<\frac{w(ck_b-k_m)}{ck_b^I-k_m-ck_b^{II}}, 
\end{equation}
which shows that the  limiting size, $R_{max}$, grows linearly with $w$. From equation (\ref{width}) we can deduce the sensitivity of the limiting size to the diffusivity,  
$R_{max}\sim D^{\frac{1}{2}}$,  and to the nutrient consumption rate, $R_{max}\sim k^{-\frac{1}{2}}$. 
A faster diffusivity will result in larger ooids and a faster nutrient consumption rate will result in smaller ooids.

Ooids typically contain concentric layers, or laminations, around a nucleus. 
The laminations may be characteristic of seasonal growth cycles,  
if there was seasonal variability in the nutrient concentration levels, 
or longer term environmental events, depending on the age of the laminations. 
Data from modern ooids~\cite{B2015} suggests that the age of ooids scales linearly with mass.
Assuming constant density then age scales linearly with volume or $R^3$. 
This means that the radius $R$ scales with time  as $t^{1/3}$.
Of course this scaling cannot continue at $t^{1/3}$ or ooids would become arbitrarily large.

The different stages of growth as described by the model ooid are depicted in Fig.~\ref{cartoon}.

\subsubsection*{Spacing of laminations and comparison with experimentally grown ooids.} 

Our mathematical model for nutrient limited ooid growth can be used to provide information on the spacing in laminations, 
by sampling the solution of the mineralisation process at constant time intervals. The radially symmetric assumption results in smooth laminations. 
In order to study the roughness of laminations, a different type of model, such as involving the 
radial version of the Kardar-Parisi-Zhang equation~\cite{BHW1998}, could be employed. 
One test of our model would be whether the spacing  between laminations, in a controlled environment,
match the spacings predicted by the model. 
Here we suppose that the mineralisation is being produced by a non-replenishing source of necrotising microbes in region I.
In this case $k_b=0$ and the model equations reduce to (\ref{eqn1})-(\ref{eqn3}).
Suppose that $R$ is the limiting radius before the onset of mineralisation and this radius is reached at time $\tau$.
We can integrate equation (\ref{eqn2}) from $\tau$ to $t$, with $V_b^I(\tau)=\frac{1}{c}\frac{4}{3}\pi R^3$ to obtain
\begin{equation}
V_b^{I}(t)=\frac{4\pi R^3}{3c}\exp(-k_b^I(t-\tau))\,.
\end{equation}
Substituting this result into (\ref{eqn1}) and integrating from $\tau$ to $t$, with $V_m(\tau)=0$ gives 
\begin{equation}
V_m(t)=\frac{4k_m\pi R^3}{3k_b^Ic}(1-\exp(-k_b^I(t-\tau))
\end{equation}
and thus the radius of the mineralisation front is given by
\begin{equation}
R_m(t)=\left(\frac{k_m R^3}{k_b^Ic}(1-\exp(-k_b^I(t-\tau)))\right)^{1/3}\,.
\end{equation}
Without loss of generality we can set $\frac{k_m R^3}{k_b^Ic}=1$ because this simply scales the size. 
Similarly we set $k_b^I=1$ because this scales the time and we can define a dimensionless time scaled by $\tau$ to obtain 
\begin{equation}
R_m(t)=(1-\exp(-(t-1)))^{1/3}, \quad t>1 
\label{result}
\end{equation}
for the radius of the mineralisation front.

The result (\ref{result}) can also be obtained in a different way. 
In order for the overall size to remain constant we need to have the microbes in the outer proliferating layer replacing
those taken up by mineralisation or decay in the inner layer.
This effect can be included directly by writing $V_b^I(t)=c(V-V_m(t))$, 
where $V$ is the fixed volume equal to that at time $\tau$.
This is saying the volume of microbes in the inner layer is being reduced as the mineralisation takes up more volume.
We would then have the mineralisation rate equation, with the rate proportional to the volume of microbes,
\begin{equation}
\frac{dV_m}{dt}=k_m c(V-V_m(t)).
\end{equation}
Solving this equation gives precisely the same dimensionless result (\ref{result}).

Initially the growth rate is seen from equation (\ref{result}) to scale as $t^{1/3}$, 
which is the characteristic growth rate in the mathematical theory of Ostwald ripening~\cite{O1897,K1975}.
As already remarked above, such a growth rate is unsustainable, as ooids are observed to be size-limited.
The result of sampling equation (\ref{result}) at constant time intervals, over long times, is shown in Fig.~\ref{figcomparison}b. 
The characteristic features are a large central region surrounded by concentric laminations whose spacing decreases over time, 
resulting in a size-limited growth.

Departures between the width and spacing of laminations of actual ooids and the laminations predicted by our model, in controlled conditions, 
could provide evidence for different environmental conditions, with different abundances of nutrients, over time in the place of formation that might 
contrast to the conditions operating at the site of deposition. 
We can now sample the result (\ref{result}) to show the position of the mineralisation front at equal time intervals.
Apart from the overall scale of the pattern, there is a single parameter given by the constant time sampling interval $\Delta t$.
The lamination rings are then located at
\begin{equation}
R_m(j)=\left(1-\exp(-(j\Delta t))\right)^{1/3}\quad j=1,2,\ldots.
\end{equation}
Fig.~\ref{figrings} shows an indicative fit of the laminations from periodic time sampling to the laminations on an  experimentally grown ooid
in a controlled environment~\cite{BKP2006}. 
The fit to the laminations is remarkable given that the only parameters to fit are the overall magnification
of the pattern, and the constant time sampling interval.

\subsubsection*{Laminations with replenishment of microbes.} 

In the model laminations described above, the microbes in region I are dying out, without replenishment, and mineralisation is occurring in this region. 
We now consider the more general case with replenishment as biotic material from region II transitions to region I.
This is described by the model equations (\ref{eqn1k})-(\ref{eqn3k}) with $k_b > 0$.
A more general version of the model taking into account that region I expands as the biofilm expands is defined by these equations.  
It is possible to solve this more general model but for simplification it suffices to consider the special case when $V_b^{II}=V_0$ is a constant. 
In this case region I is continually being supplied with microbial material from region II. 
Suppose that mineralisation begins at time $\tau$, in this case with an initial volume  $V_b^I(\tau)$ in region I, and $V_b^{II}(\tau)=V_0$ in region II. 
Similar analysis of this case leads to the more general result 
\begin{equation}
R_m(t)=\left(1+\alpha(t-1)-\exp(-(t-1))\right)^{1/3},\quad t>1, 
\end{equation}
where
\begin{equation}
\alpha=\frac{k_b^2V_0}{k_b^IV_b^I(\tau)-k_b V_0}.
\end{equation}
When $\alpha=0$ this recovers the previous result (\ref{result}) for $R_m(t)$. 
The temporal scaling behaviour $R_m(t)\sim t^{1/3}$ for $t\gtrapprox1$ holds for all $\alpha$. 
Moreover, the pattern of laminations does not differ significantly when $\alpha\neq 0$.

\subsection*{Diagenetic modification of the ideal ooid structure.} 

It has been clearly shown that ooids are very susceptible to authigenic and diagenetic change (see, e.g., Fig.~1.2 in ref.\cite{TWD1990}). 
In reality the ideal ooid structure may be complicated by syngenetic aggrading neomorphism of the mineral phase\cite{B1971} 
with Ostwald ripening~\cite{MS2011,WC2014} being the likely driver of this recrystallisation (Figs \ref{figvarious}d and \ref{figcomparison}a). 
Organic matter and other impurities are rejected by the growing crystallites and form a boundary layer ahead of the growth front\cite{MJ2009}. 
Davies {\em et al.}~\cite{DBF1978} have suggested a modified version of Sorby's theory in which the evolving ooid alternates between 
`suspension' and `resting' growth phases.  
We suggest instead that ooids are created by the interaction of the mineralising results of biofilm accretion, 
superimposed by the effects of syngenetic mineral growth. 
Here we illustrate an example of how the original ooid fabric may be modified by diagenetic changes. 
Kalkowsky\cite{K1908} described details of several diagenetic structures that have modified the original concentric ooid. 
Fig.~\ref{figheeseberg} shows ooids of approximately 0.5 to 1cm in diameter from the Triassic Rogenstein of the Heeseberg Quarry\cite{P1999}. 
These were originally spherically concentric ooids, as described by our model. 
The apparent branching is probably caused by the effects of syngenetic and diagenetic mineralisation superimposed on the primary concentric structures. 
From the observed patterns it is tempting to suggest that the `growth' of the branches was controlled from the outset 
by the radial version\cite{BHW1998} of the KPZ equation~\cite{KPZ}. 
Branching can occur when anisotropic effects (e.g., noise or amplification of a small bump through an instability mechanism) compete with surface tension. 
In fact a model has been developed for conical stromatolites in which diffusive gradients, 
extending over the thickness of an overlying microbial mat, causes mineral precipitation to be faster in regions of high curvature\cite{X2013}. 
The mathematical model is similar to that for growth of stalactites\cite{SBG2005} and icicles\cite{SBG2006}.

If the effects of syngenetic mineral growth dominates the mineralising results of biofilm accretion,  
for example as nutrient availability becomes limited and surface tension is reduced, 
or during later diagenetic recrystallisation, structural complexity is superimposed on the pattern predicted by our model (Fig.~\ref{figheeseberg}), 
ultimately creating ooids exhibiting the {\em spindelstruktur} and {\em kegelstruktur} described by Kalkowsky\cite{K1908} (tafel iv, Figs 2 and 3 therein) 
with surface protruberences giving rise to their description as cerebroid ooids\cite{Carozzi}. 
We suggest that {\em spindelstruktur} and {\em kegelstruktur} provide evidence  for the existence  of competing  processes of ``Greenspan" biofilm accretion  
and ``Ostwald" mineralisation operating simultaneously in  ooid  genesis.   
For  ooids  we  assume  radial  symmetry  when  the  influence of biofilm accretion dominates.      
A diagnostic characteristic of an ooid is that any irregularities in the nucleus are damped progressively in successive layers within the cortex 
until a spherical form is established. 
This observed behaviour can be explained by the fact that radially symmetric growth will occur when the surface tension is dominant in the growth process. 
In simulations involving the radial KPZ equation, irregular initial shapes grow to be either circular or spherical, depending on the dimensionality\cite{BHW1998}. 
When surface tension dominates in the KPZ equation there is only diffusion.
Mineralisation initially takes place within these concentric layers, but diffusion limited mineral growth gradually develops the radial {\em spindelstruktur}  
(Fig.~\ref{figheeseberg}a) and increasing mineral growth in the areas between will eventually lead to the development of {\em kegelstruktur} (Fig.~\ref{figheeseberg}b) 
forming outward projecting bumps that will grow faster through diffusion limited growth.

\section*{Discussion}

Despite extensive research over more than a century there is still a lack of conclusive information on the genesis of ooids \cite{B2012,F1977}.  
One limitation is that, although  highly sophisticated geobiological analyses are  now being made, 
they are  undertaken on samples  collected using 19th century techniques (e.g., refs\cite{DSEODSA2015,OReilly2017}).
As pointed out by Fabricius \cite{F1977}, it would be a mistake to assume that a similar process has formed all concentrically laminated grains.  
Clearly the structure of the typical ooid does not reflect successive surface accretion of carbonate mud on the surface of a rolling grain, 
as suggested by Sorby, but rather provides a detailed archive of organo-sedimentary concentric accumulation. 
Informed analysis of this archive may elucidate the detailed history of the growth of the ooids~\cite{L2017}. 
Such analysis is necessarily complicated by the task of discriminating whether growth corresponds to the model outlined in this paper, 
whether the original fabric has been overprinted by subsequent mineralisation, or even whether an entirely different process is 
responsible for the concentric structure  (e.g., Fig.~6 in ref.\cite{YZ2014}). 
Although these discussions generally concentrate on carbonate ooids, our model will equally apply to ooids mineralised  
by chamosite, stevensite, phosphate or other mineral phases. 

\begin{figure}[ht]
\centering
\includegraphics[width=\linewidth]{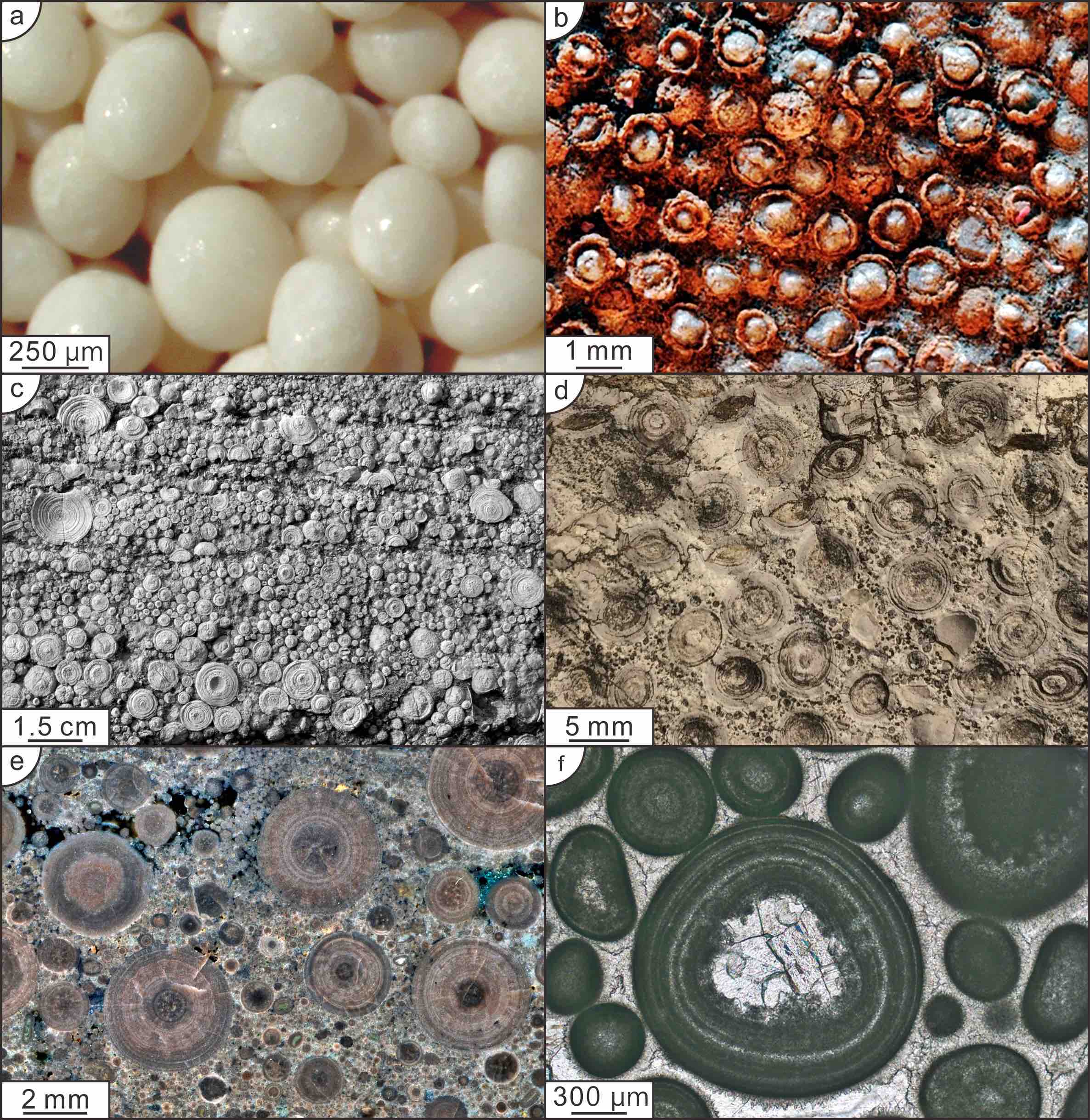}
\caption{
\textbf{Variations of size and sorting of ooids.} 
(\textbf{a}) Typical Bahamian ooids from a sand wave at Joulter's Cay, Bahamas, 
showing sorting and polished surfaces from grain collisions during transport from point of formation (Image courtesy Mark Wilson). 
(\textbf{b}) Comparable perfectly sorted ooids in an Archean (2.72 Ga) oolite in the Tumbiana Formation, Pilbara, Western Australia. 
(\textbf{c}) Cross section of Rogenstein ooids (Triassic) in a block mined from a quarry described by Br\"uckmann in 1721 showing giant ooids, 
poor sorting and cross sections with concentrically laminated corteces (Triassic, Kirchstra\ss e, Barneberg, Germany).  
(\textbf{d}) Poorly sorted ooids showing typical cross sections of core and cortex with alternatively dark- and light-laminae, 
note tendency to a maximum diameter of $\sim4$mm (Middle Cambrian, Longmen, China). 
(\textbf{e}) Photomicrograph of thin section of typical Triassic Rogenstein ooids showing concentric layers and {\em kegelstruktur} and {\em spindelstruktur} overprint. 
Heeseberg, Germany. 
(\textbf{f}) Thin section of cross sections of ooids (Triassic, Lichuan, China) showing core, concentric layers and diagenetic overprint of later crystal accumulation.
}
\label{figvarious}
\end{figure}

\begin{figure}[ht]
\centering
\includegraphics[width=\linewidth]{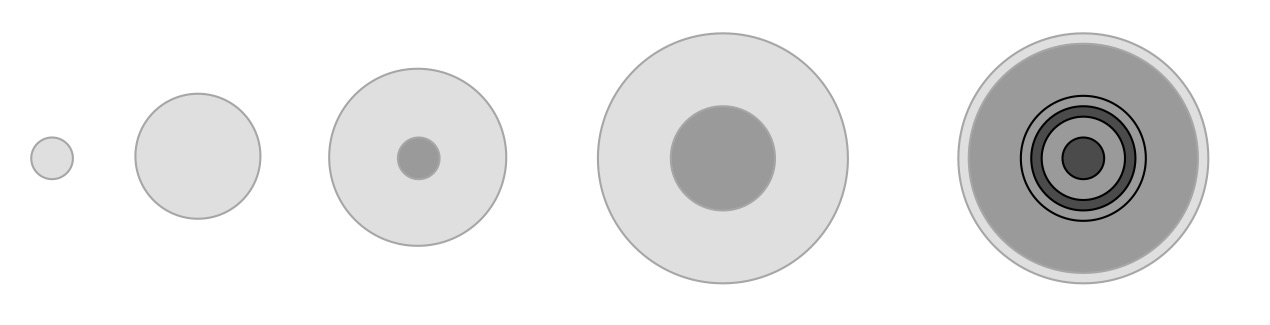}
\caption{
\textbf{Stages of growth described by the model ooid.} 
In the initial stage a biofilm contains microbes that are supplied by diffusing nutrients. In the intermediate stage nutrients are consumed by microbes in the outer region 
before they can diffuse into the inner region. In the later stage microbes die in the inner zone and mineralisation occurs. The outer region limits to a constant width zone.}
\label{cartoon}
\end{figure}

\begin{figure}[ht]
\centering
\includegraphics[width=\linewidth]{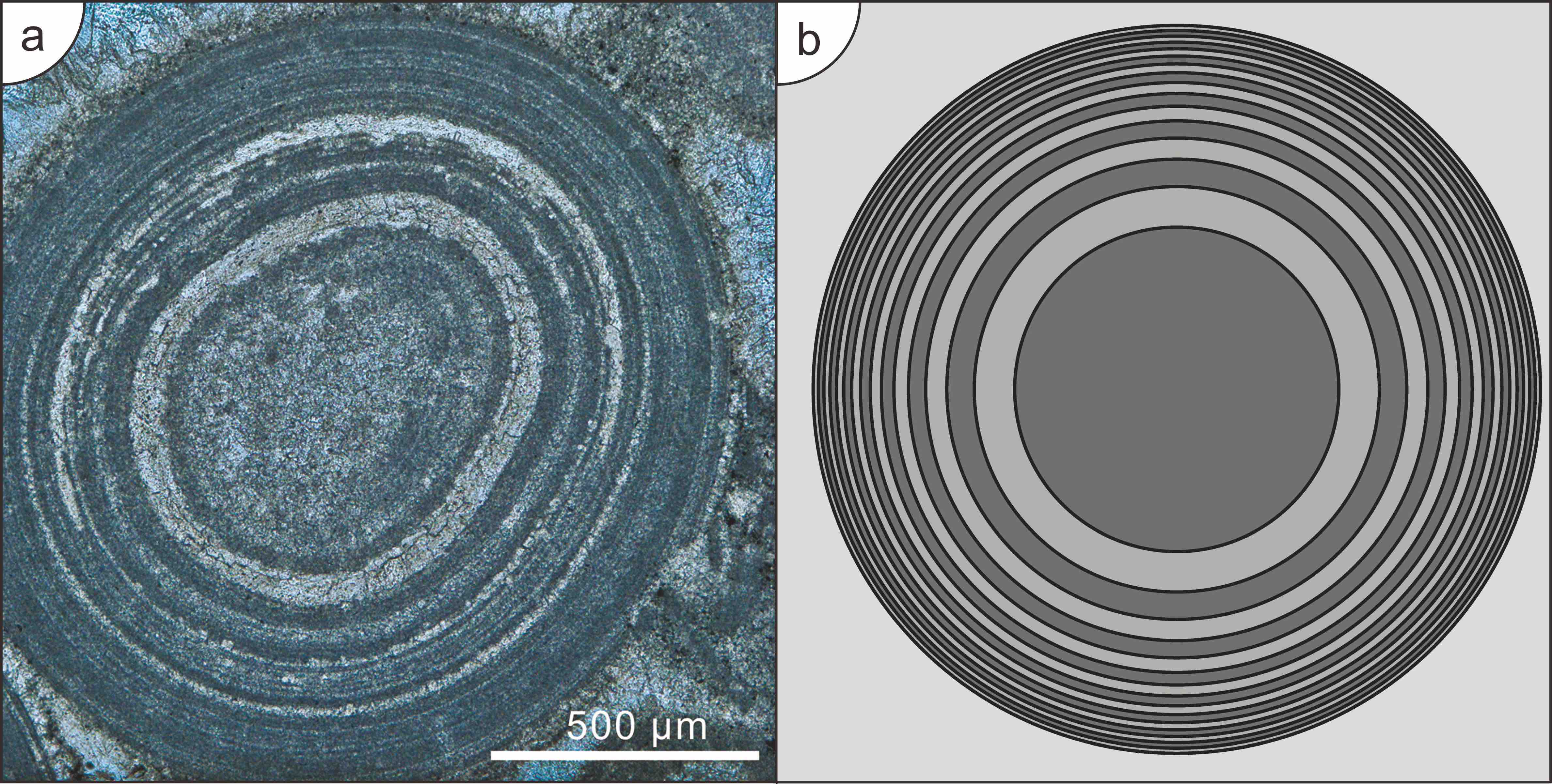}
\caption{
\textbf{Comparison of the model ooid with an actual Triassic ooid.} 
(\textbf{a}) Thin section from the Lower Triassic of Pingguo, China. 
(\textbf{b}) Laminations at uniform time intervals under constant parameter growth conditions in the model ooid.}  
\label{figcomparison}
\end{figure}

\begin{figure}[ht]
\centering
\includegraphics[width=\linewidth]{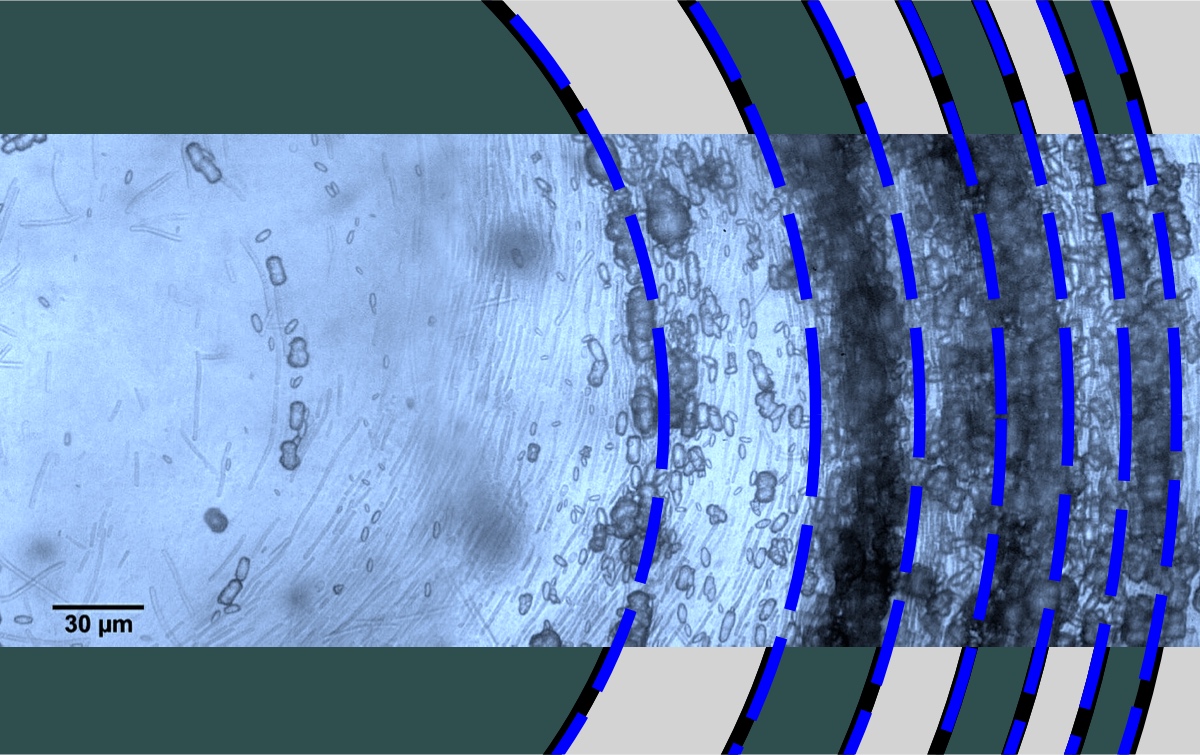}
\caption{
\textbf{Comparison between theory and experiment.} 
Representative comparison between the laminations at uniform time intervals under constant parameter growth conditions in the model ooid
and the laminations in the laboratory ooid grown by Brehm, Krumbein and Palinska\cite{BKP2006}. Background image supplied by 
Katarzyna Palinska.} 
\label{figrings}
\end{figure}

\begin{figure}[t!]
\centering
\includegraphics[width=\linewidth]{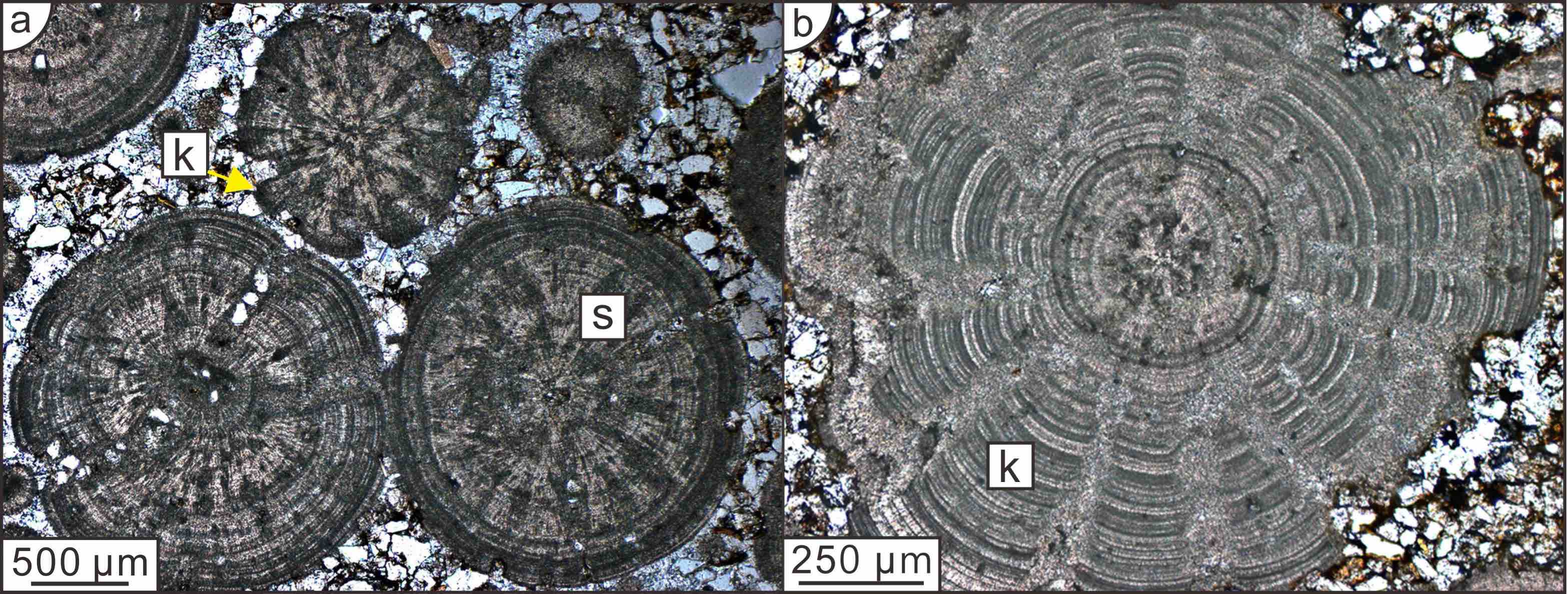} 
\caption{
\textbf{Diagenetic modification of original ooid structure.} 
(\textbf{a}) Rogenstein ooid from Heeseberg showing early development of {\em spindelstruktur} (s) as a result of syngenetic mineral change in the cortex. 
Spindles create zones of weakness that demarcate the boundaries of the components of the incipient {\em kegelstruktur} (k). 
(\textbf{b}) Ooid from the Rogenstein of Heeseberg showing fully developed {\em kegelstruktur} (k). Note convex nature of originally concentric laminae within each 
kegel indicating differentially faster growth at the apex of each protruberence.}  
\label{figheeseberg}
\end{figure}

One consequence of the model presented here is that the resulting ooids would grow to a limiting maximum size.  
Traditionally ooids have been arbitrarily regarded as having a diameter of 2mm or less (Figs~\ref{figvarious}a, \ref{figvarious}b), 
but this was never specified originally\cite{K1908,B2012}.   
It is clear that ooids can attain much larger sizes (Fig.~\ref{figvarious}).  
Such ``giant" ooids have been explained in terms of Sorby's model by formation under conditions of high current velocity\cite{SG1993}, 
but Fig.~\ref{figvarious}c clearly shows an assemblage of poorly sorted ``giant" ooids that clearly do not support the evidence of accumulation by a high velocity current. 
From the perspective of the model presented in this paper the only difference between ``giant" ooids and conventional ooids is that the former 
grow under more favourable conditions reflected in the model parameters (e.g., availability of biomass and nutrients).  
The occurrence of large accumulations of well sorted ooids in units showing 
current-generated depositional structures (Figs~\ref{figvarious}a, \ref{figvarious}b) raises the question as to whether these ooids have been reworked and concentrated together, 
as Br\"uckmann\cite{B2012} suggested in 1721 when he wrote ``a global wind (Aeolus macrocosmicus), 
ruling during the flood and stirring the waters, drove the eggs flowing in the waters of the flood to (concentrate in) certain places". 
Evidence for this reworking could include abraided, polished grain surfaces (Fig.~\ref{figvarious}a) and evidence of microbial boring within the ooids\cite{B2015}.
Our model thus  provides a potential  key  to understanding the genetic information preserved in the internal structures of naturally  
occurring ooids that might reflect environmental conditions in complete contrast to those operating at the site of final deposition, 
for example, the ooids of Kalij el-Arab \cite{Anwar}.

\section*{Acknowledgements}

We thank Mark A. Wilson for permission to reproduce his photograph as Fig.~1a and Katarzyna Palinska 
for providing the background image for Fig.~3. 
We also thank Christian Renggli and Martin Schaefer for assistance with the translation of Kalkowsky's paper\cite{K1908}  
and Nigel Goldenfeld for a discussion on the theory of Ostwald ripening. 
The work of F.L. has been supported by the National Natural Science Foundation of China (Grant No.~41502115). 
The work of M.T.B. has been supported by the 1000 Talents Program of China.

\section*{Author Contributions}

J.P. performed field and lab studies of Triassic ooids in Germany. 
F.L. performed field and lab studies of Triassic ooids in China. 
R.V.B. performed field and lab studies of Triassic ooids from Germany and China and modern ooids in Bahamas and Shark Bay.
B.I.H. and M.T.B. performed the calculations and analysis of the mathematical model. 
All authors contributed to the development of the concepts and writing of the paper.

\section*{Additional Information}

\textbf{Competing financial interests:} The authors declare no competing financial interests.

\vskip 8mm

\begin{table}[h]
\centering
\begin{tabular}{|c|l|}
\hline
$V_m$ & volume of mineralised material in the biofilm \\
\hline
$V_b^I$ & volume of biotic material in the inner region of the biofilm \\
\hline
$V_b^{II}$ & volume of biotic material in the outer layer of the biofilm \\
\hline
$c_I$ & threshold nutrient concentration needed for microbes to survive \\
\hline
$c_{II}$ & background nutrient concentration  \\
\hline
$c$ & scale factor relating the volume of biofilm to the volume of microbes in the biofilm \\
\hline
$k$ & rate of transition of microbes from the outer layer to the inner zone\\
\hline
$k_m$ & constant rate of conversion of biotic material to mineralisation in the inner zone \\
\hline
$k_b$ & constant per volume rate that microbes from the outer layer transition into the inner zone\\
\hline
$k_b^I$ & constant per volume rate of decay of microbes in the inner zone  \\
\hline
$k_b^{II}$ & constant per volume rate of decay of microbes in the outer layer  \\
\hline
$D$ & diffusivity of nutrients  \\
\hline
$R$ & limiting radius of the biofilm  \\
\hline
$w$ & limiting width of the outer layer  \\
\hline
$t$ & time  \\
\hline
\end{tabular}
\caption{The meanings of variables and parameters in the model equations.}
\label{tabvars}
\end{table}


\begin{thebibliography}{9}

\bibitem{Flugel1978} 
Fl\"ugel, E. 
{\em Mikrofazielle Untersuchungsmethoden von Kalken} Springer-Verlag (1978).

\bibitem{Simone1980}
Simone, L. 
Ooids: a review. 
{\em Earth-Science Reviews} {\bf 16}, 319-355 (1980).

\bibitem{Richter1983}
Richter, D. K. 
Calcareous ooids: a synopsis. 
In {\em Coated grains} Springer Berlin Heidelberg (1983) pp 71-99. 


\bibitem{B2012}
Burne, R. V., Eade, J. C. \& Paul J. 
The Natural History of Ooliths: Franz Ernst Br\"uckmann's treatise of 1721 and its significance for the understanding of oolites. 
{\em Hallesches Jb. Geowiss.}  {\bf 35}, 93-114 (2012).

\bibitem{S1879}
Sorby, H. C. 
The structure and origin of limestones. 
{\em Proc. Geol. Soc. London} {\bf 35}, 56-94 (1879).
 
\bibitem{B1971}
Bathurst, R. G. 
Carbonate sediments and their diagenesis.  
{\em Developments in Sedimentology}, Vol. 12, Elsevier (1971).

\bibitem{SF1972} 
Suess, E. \& F\"utterer, D.  
Aragonitic ooids: experimental precipitation from seawater in the presence of humic acid. 
{\em Sedimentology} {\bf 19}, 129-139 (1972).

\bibitem{S1970} 
Shearman, D. J., Twyman, J. \& Karimi, M. Z. 
The genesis and diagenesis of oolites. 
{\em Proceedings of the Geologists' Association} {\bf 81}, 561-575 (1970).  

\bibitem{Trower2017}
Trower, E. J., Lamb, M. P.  \& Fischer, W. W. 
Experimental evidence that ooid size reflects a dynamic equilibrium between rapid precipitation and abrasion rates. 
{\em Earth and Planetary Science Letters} {\bf 468}, 112-118 (2017).


\bibitem{SG1993}
Sumner, D. Y. \& Grotzinger, J. P. 
Numerical modeling of ooid size and the problem of Neoproterozoic giant ooids. 
{\em Journal of Sedimentary Research} {\bf 63}, 974-982 (1993).

\bibitem{S2003}
Stewart, P. S.  
Diffusion in biofilms. 
{\em Journal of Bacteriology} {\bf 185}, 1485-1491 (2003).

\bibitem{L2017} 
Li, F. {\it et al.} 
Paleo-seawater REE compositions and microbial signatures preserved in laminae of Lower Triassic ooids. 
{\em Palaeogeography, Palaeoclimatology, Palaeoecology} {\bf 486},  96-107 (2017).

\bibitem{BKP2006}
Brehm, U., Krumbein, W. E.  \& Palinska, K. A. 
Biomicrospheres generate ooids in the laboratory.
{\em Geomicrobiol. J.} {\bf 23}, 545-550 (2006).

\bibitem{Diaz2017} 
Diaz, M. R. {\em et al.}
Microbially mediated organomineralization in the formation of ooids. 
{\em Geology} {\bf 45} 771-774 (2017).

\bibitem{G1972}
Greenspan, H. P. 
Models for the growth of a solid tumor by diffusion. 
{\em Studies in Applied Mathematics} {\bf 51}, 317-340 (1972).

\bibitem{B2015} 
Beaupr\'e, S. R., Roberts, M. L., Burton, J. R. \& Summons, R. E. 
Rapid, high-resolution 14 C chronology of ooids. 
{\em Geochimica et Cosmochimica Acta} {\bf 159}, 126-138 (2015).

\bibitem{BHW1998}
Batchelor, M. T.,   Henry, B. I. \& Watt, S. D. 
Continuum model for radial interface growth. 
{\em Physica A} {\bf 260}, 11-19 (1998).

\bibitem{O1897}
Ostwald, W. 
Studien \"uber die Bildung und Umwandlung fester K\"orper.
{\em Z. Phys. Chem} {\bf 22}, 289-330 (1897).  

\bibitem{K1975}
Kahlweit, M. 
Ostwald ripening of precipitates. 
{\em Advances in Colloid and Interface Science} {\bf  5}, 1-35 (1975).

\bibitem{TWD1990}
Tucker, M. E.  \& Wright, V. P. 
{\em Carbonate Sedimentology}, Blackwell (Oxford), 482 pages (1990).

\bibitem{MS2011}
Melim, L. A. \& Spilde, M. N. 
Rapid growth and recrystallization of cave pearls in an underground limestone mine. 
{\em Journal of Sedimentary Research} {\bf 81}, 775-786 (2011).

\bibitem{WC2014}
Li, S., Wang, Z.-J.  \& Chang, T.-T.  
Temperature Oscillation Modulated Self-Assembly of Periodic Concentric Layered Magnesium Carbonate Microparticles. 
{\em PloS One} {\bf 9} e88648 (2014).

\bibitem{MJ2009}
Meakin, P. \& Jamtveit, B. 
Geological pattern formation by growth and dissolution in aqueous systems.
{\em Proc. R. Soc. A} {\bf 466}, 659-694 (2010).

\bibitem{DBF1978}
Davies, P. J., Bubela, B.  \& Ferguson, J.  
The formation of ooids. 
{\em Sedimentology} {\bf 25}, 703-730 (1978).

\bibitem{K1908}
Kalkowsky, E. 
Oolith und Stromatolith im norddeutschen Buntsandstein. 
{\em Zeitschrift der Deutschen Geologischen Gesellschaft} 
{\bf 60}, 68-125 (1908).


\bibitem{P1999}
Paul, J. 
Oolithe und Stromatolithen im Unteren Buntsandstein,  
III.6 in Trias, eine ganz andere Welt: Mitteleuropa im fr\"uhen Erdmittelalter. Edited by Norbert Hauschke. Pfeil, pp 263-270 (1999).

\bibitem{KPZ} 
Kardar, M., Parisi, G. \& Zhang, Y.-C. 
Dynamic scaling of growing interfaces. 
{\em Phys. Rev. Lett.} \textbf{56}, 889-892 (1986).

\bibitem{X2013}
Petroff, A. P., Beukes, N. J., Rothman, D. H. \& Bosak, T. 
Biofilm growth and fossil form. 
Phys. Rev. X {\bf 3}, 014012 (2013).

\bibitem{SBG2005}
Short, M. B., Baygents, J. C.  \& Goldstein,  R. E.  
Stalactite growth as a free-boundary problem.
{\em Phys. Fluids} \textbf{17}, 083101 (2005).

\bibitem{SBG2006}
Short, M. B., Baygents, J. C.  \& Goldstein,  R. E.  
A free-boundary theory for the shape of the ideal dripping icicle. 
{\em Phys. Fluids} \textbf{18},  083101 (2006).

\bibitem{Carozzi}
Carozzi, A. V. 
Cerebroid oolites. 
{\em Trans. Illinois State Acad. Sci.} {\bf 55}, 239-249 (1962).

\bibitem{F1977} 
Fabricius, F. H. 
Origin of marine ooids and grapestones. 
{\em Contributions to Sedimentology} {\bf 7}, 113 pages 
E. Schweizerbart'sche Verlagsbuchhandlung, Stuttgart. (1977).

\bibitem{DSEODSA2015}
Diaz, M. R.  {\it et al.} 
Geochemical evidence of microbial activity within ooids. 
{\em Sedimentology} {\bf 62}, 2090-2012 (2015).

\bibitem{OReilly2017}
O'Reilly, S. S. {\it et al.} 
Molecular biosignatures reveal common benthic microbial sources of organic matter in ooids and grapestones from Pigeon Cay, The Bahamas. 
{\em Geobiology} {\bf 15}, 112-130 (2017).

\bibitem{YZ2014}
Yec, C. C.  \& Zeng, H. C.  
Synthesis of complex nanomaterials via Ostwald ripening. 
{\em Journal of Materials Chemistry A} {\bf 2}, 4843-4851 (2014).

\bibitem{Anwar} 
Anwar, Y. M., El Askary, M. A.  \& Nasr, S. M. 
Arab's Bay oolitic carbonate sediments: bathymetric, granulometric and chemical studies. 
{\em Neues Jahrbuch F\"ur Geologie und Pal\"aontologie-Monatshefte} {\bf 10} 594-610 (1984). 




\end{thebibliography}
\end{document}